\documentclass[a4paper,10pt]{article}
\usepackage{a4}
\begin{document}
\title{Discrete spectral triples converging to dirac operators.}
\author{Alejandro Rivero\thanks{ 
Email: \tt rivero@wigner.unizar.es}}
\maketitle
\begin{abstract}

We exhibit a series of discrete spectral triples converging  
to the canonical spectral triple of a finite 
dimensional manifold. Thus we bypass the non-go theorem of
G\"okeler and Sch\"ucker. Sort of counterexample.

\end{abstract}

In \cite{connes} Connes gave the most general example of non commutative 
manifolds defined from finite spectral 
triples, and such discrete manifolds were classified 
independently by Krajewsky \cite{kr1,kr2} and Paschke and 
Sitarz \cite{PS} the 
same year. But almost the same time, G\"okeler and 
Sch\"ucker \cite{GS} proved that
 the naive discretizations of Dirac Operators 
were not able to fulfil the axioms of non-commutative 
manifolds. This caused a fading in the attempts to 
illuminate NCG from the point of view of lattice theories. 
Later attempts to show NCG operations in 
lattices have preferred to avoid the full axiomatic 
required by K-theory and Reality \footnote{In NCG, Reality refers to 
Atiyah' work. I can not resist to quote J.P. May: ``Atiyah introduced
Real K-theory KR, which must not be confused with real K-theory KO. In the
paper, real vector bundles mean one thing over {\it real spaces} and
another thing over {\it spaces}, which has bedeviled readers ever since: we
distinguish Real from real, never starting a sentence with either''. It is
a pity that Atiyah's original does not follow this suggestion.}.
But the subsequent works on lattices, from the successful show of 
L\"uscher with the index theorem, to the insistent studies of Jian
Dai or M. Requardt, to name some ones, should invite us to try to develop
the full geometric set-up.

In \cite{GS}, the non-go principles also ruled out some non-naive 
formulations. At that time,
inspired on some previous 
speculations, I risked to suggest privately the use
 of a doubling of the representation of the algebra, thus a 
function $A(x)$, $x$ in one dimension,  
could be discretized to an operator $A(i)$ with representation:
$$
A |v>  =\pmatrix{ \ddots  & & &  &  &   \cr
                          &  A_{i-2} &&&&       \cr
                          &&    A_{i-1} &&&     \cr
                          &&&      A_{i-1} &&   \cr
                          &&&&         A_i   &  \cr
                          &&&&&         \ddots    }
\pmatrix{\vdots \cr v_{i-1}^- \cr v_{i-1}^+ \cr
                 v_i^+  \cr v_i^- \cr \vdots}
$$
Regretfully this doubling is not a non-commutative 
manifold, as a check of \cite{kr1} or \cite{PS} can show.  So it 
was not a good idea. Now, it happens that the axioms can
 be met if we opt by a tripling, i.e. a representation of 
$A(i)$ as:
$$
A |v>  =\pmatrix{ \ddots  & & &  &     \cr
                          &    A_{i-1} &&&     \cr
                          &&      A_{i} &&   \cr
                          &&&         A_{i+1}   &  \cr
                          &&&&          \ddots    }
\pmatrix{\vdots \cr v_{i}^- \cr v_{i}^0 \cr
                 v_i^+   \cr \vdots}
$$
This representation corresponds, in Connes example, to 
the commutative discrete spectral triple with 
intersection matrix\footnote{I want to thank the colective aid of the
people of usenet newsgroup 
 {\tt sci.math}, time ago, to understand this family of matrices}:
$$
q=\pmatrix{  -1 & 1 & & 0 & b  \cr 
              1 & -1 & 1 & & 0  \cr 
           & 1 & \ddots & 1 &   \cr 
              0 & & 1 & -1 & 1 \cr 
              b & 0 & & 1 & -1 \cr}
$$
where $b=1$ to get a discrete circle, and $b=0$ to get a segment.

There were in principle two reasons to disregard this 
triple. On one hand, the intersection matrix seems to 
be degenerated. It is not. On the other hand, the 3x3 
matrix over each point seems not to be able to 
reproduce a set of Dirac (or at least Pauli) matrices 
in the limit. It can.

The degeneracy of the matrix is accidental and depends 
of the dimension of the matrix, this is, of the 
number of points we are using in the corresponding 
discrete lattice. We can take determinant of q and we see\footnote{Of course
there are two null eigenvectors for the circle but only one for
the segment} that the 
sequence is, for the circle,
$$1,-3,4,-3,1,0,1,-3,4,-3,1,0,1,-3,\dots$$
where the first 0 happens for n=6 and with periodicity six,
while for the the segment it happens
first for n=2 and then with periodicity 3:
$$-  , 0,1,-1,0,1,-1,0,1,-1,0,1,\dots$$
so in both cases is it possible to take a non degenerate 
subsequence. Moreover, one can cast some doubt 
about the causes of the failure of the intersection form: 
the axiomatic asks for Poincare duality in the 
K-theory, and additional properties must be requested to 
the Chern character in order to get the intersection 
form as commonly is calculated. Perhaps a fine-tuning 
in the Dirac operator could give us a character 
wiping away the degeneracy of the intersection. 

As for the second point, one can observe that the 
operator $[D,A]$ has always a null space of dimension one 
third of the total. In the infinite limit, this 
nullspace becomes independent of A and we recover the usual 
expression $i \sigma_1 \partial_x$. 

Lets build more explicitly the spectral triple, following the instructions
of \cite{connes}. The hilbert space is the sum of subspaces $\oplus H_{ij}$,
the dimension of each subespace being given by the absolute value of
the element $q_{ij}$. There is a chirality operator $\gamma$ that acts
on $H_{ij}$ as $\pm 1$, following the sign of $q_{ij}$. There is also
a involution $J$ with action $J |e_{ij}>  =  |\bar e_{ji}>$

The action of A in each $H_{ij}$ is multiplication times $A(i)$. This is the
kind of matrix mentioned above. Note that $J A J^{-1}$ builds the opposite
algebra $A^O$.

Finally, the dirac operator is given by elements $m_{ij,kl}$ having some
restrictions: first, there are different of zero only if $i=k$ or
$j=l$. Second, they are different of zero only when the pairs $ij$ and
$kl$ show different sign (ie chirality) in the matrix $q$. Third, they
are restricted by the symmetries $m_{ij,kl}=\bar m_{kl,ij}$ and
$m_{ij,kl}=\bar m_{ji,lk}$. With these symmetries, is is easy
to check that that the differential [D,A] is a matrix of diagonal
boxes, acting in the subespaces $H_{-,l}$ and codifying the forward
and backward derivatives, $a'_-, a'_+$ at the point $l$.

It is instructive to see the nullspace of the differential 
with some 
detail. Thus explicitly, for a given point $l$, the operator $[D,A]$ on the 
subspace $H_{l-1,l} \oplus H_{l,l} \oplus H_{l+1,l}$ is:
$$[D,A]|> \;  = \pmatrix{
  0  &     (a_l-a_{l-1}) m_{l-1\, l, l l} &  0   \cr
 (a_{l-1}-a_l) m^*_{l-1\, l, l l} & 0   &   (a_{l+1}-a_l) m_{l l, l+1\, l} \cr
  0 &  (a_l-a_{l+1}) m^*_{l l, l+1\, l} & 0 }
\pmatrix{e_{l-1,l} \cr e_{ll} \cr
                 e_{l+1 \, l}   }
= 
$$ 
$$
=i   \pmatrix{ 
                     0   & a'_-  &  0    \cr
                    a'_-    & 0   &  a'_+   \cr
                     0   &  a'_+  & 0      }
\pmatrix{e_{l-1,l} \cr e_{ll} \cr
                 e_{l+1 \, l}   }
$$
(note that $A^o$ acts here diagonnally, $A(l) {\bf 1}_3$, fullfilling the
first order condition of the differential calculus)

The nullvector comes from a lineal combination 
of the forward and backward derivatives, $|\Omega> = a'_+ |e_{l-1,l}-
 a'_- |e_{l+1,l}>$.
We can rotate this vector to the basis to see 
explicitly the zero; get vectors $(0,\sqrt{{a'}_-^2+{a'}_+^2},0)$ and
 $(-a'_-,0,a'_+)$ and then $[D,A]$ is:
$$
    \pmatrix{ 
                     0   & 0  &  0    \cr
                     0    & 0   &   i \sqrt{(a'_-)^2+(a'_+)^2}   \cr
                     0   &  i \sqrt{(a'_-)^2+(a'_+)^2}   & 0      }
\pmatrix{e_\alpha \cr e_{ll} \cr
                 e_\beta   }
$$

In the limit, forward and backward derivatives
 always coincide, so the (normalized) change is the same for each box
and can be done simultaneously 
over the total summed vectors 
$$E^-= \oplus_l e_{l-1,l},\;
E^+= \oplus_l e_{l+1,l},\;E^0=\oplus_l e_{l,l}$$ 
  	
The nullspace $E^- - E^+$ is completely contained in the positive 
chirality sector; which cures a non-problem; we had 
built a hilbert space where the positive chirality 
had double number of degrees of freedom that the 
negative one. In the commutative limit, bot chiralities are equal.
The chirality operator, $\gamma$, is then other Pauli matrix.

So we see that it is possible to build a one-dimensional manifold
 as the limit of zero dimensional spectral 
triples. The discrete spectral triple is  0-dimensional. So we 
can apply the rule for the product of even-dimensional spectral triples 
to get  0-dimensional triples having as limit higher dimensional manifolds. 
Just to check as it works, we can multiply the basic spectral triple 
by itself. The algebra is now the direct product of two copies of $A$, and the
same rule applies to the hilbert spaces $H$ and, at least in the
even case, to the operator $J$. As for the dirac operator, in this
case (and to get a good rule for $D^2$) it is defined to be:
$$D_2 = D \otimes 1 + \gamma \otimes D $$

So  the commutator $[D_2,A^x\otimes A^y]$
 results
$$
[D,A^x] \otimes A^y + \gamma A^x \otimes [D,A^y]
$$
getting a new Pauli matrix in the second term; the continuous
 limit give us the previous $\sigma_1$ and then the result
is $$
\lim [D_2, A\otimes A]= 
 i \sigma_1 \partial_x A^x \otimes A^y + (-\sigma_3) A^x \otimes 
   i \sigma_1 \partial_y A^y 
\equiv i (\sigma_x \partial_x + \sigma_y \partial_y)$$
not so good as expected.

It is possible to iterate this process, first doing the calculation
in the discrete, then taking the limit. As the products are always of
0-dimensional spectral triples, this method avoids the doubts\cite{vanhecke}
about which sign rule applies to the product, depending on dimension.

To end, some remarks:

a) It still seems interesting to study the action $Tr_\omega (D^{n-2})$, where
$n$ is the dimension of the triple.

b) The naive attempt, with only a doubling, merits further study. It is 
apparent that the $q$ matrix is invertible or degenerate depending of the
existence of boundary for the limit manifold. Also it needs of an even number
of points, to hide the inhomogeneity of this kind of spaces of alternating
chirality. In the lattice, should be interesting a connection to
continuum limits of antiferromagnetic states. 
 
c) The motivation to try this discretization comes from previous conjectures
about physicsl calculus. But
while physically it could exist some suggestion, mathematically one feels unsure about
the need of this duplication (and triplication) of degrees of freedom. Besides the
usual stuff on Clifford Algebras, I wonder if it could be related with the 
complications to embbed a manifold in an euclidean space... it could be a motivation
to go to the cinema this month. But if it were, it is good to remember \cite{Clarke}
that the
non-riemannian metrics need higher dimensions, going (casually) up to dimension
ninety for the (3,1) metric\footnote{To be precise, $2+46$ for compact
spaces and $2+87$ for the non-compact case. But see also \cite{greene}}.

d) It seems that there is some independence between the forward
and backward derivatives, box-to-box. If this can be manifest and if this
implies some freedom in D, which could survive as discrete space in the
continous limit, it remains as a note for further research.

\end{document}